\documentstyle[preprint,prl,aps]{revtex}

\begin{document}
\draft

\title{
Resonating Valence Bond Theory of Coupled Heisenberg Chains
}
\author{ S.R.\ White and R.M.\ Noack }
\address{
Department of Physics
University of California
Irvine, CA 92717
}
\author{ D.J.\ Scalapino }
\address{
Department of Physics
University of California
Santa Barbara, CA 93106
}
\date{\today}
\maketitle
\begin{abstract}
Using numerical results from a density matrix renormalization group
study as a guide, we develop a resonating valence bond (RVB) theory for
coupled Heisenberg chains.
We argue that
simple topological effects mandate a short-range RVB description of
systems with an even number of chains $n_c$,
with a spin gap, short-range correlations,
and confinement of topological spin defects.
Odd-$n_c$ systems have long-range RVB
ground states, no gap, and power-law correlations.
\end{abstract}


\narrowtext

The discovery of materials such as $(\text{VO})_2 \text{P}_2
\text{O}_7$ \cite{johnston} and $\text{Sr}_2 \text{Cu}_4 \text{O}_6$
\cite{takano}, which contain weakly coupled arrays of
metal-oxide-metal ladders, has stimulated interest in coupled-chain
Heisenberg, Hubbard, and $t$-$J$ systems.
A number of studies \cite{dagotto,strong,barnes,rice1,parola,rice2}
have provided strong evidence that at
half-filling the two-chain ladder systems are spin liquids, with a
spin-gap and finite spin-spin correlation length.
It has been proposed that the system
can be described in terms of a short-range resonating valence bond
(RVB) picture
\cite{rice1,rice2}, and mean field treatments with Gutzwiller
renormalization of the matrix elements have supported this conclusion.
It has also been suggested that in contrast, systems with an odd number
of chains $n_c$ are gapless\cite{parola,rice2}.

In this paper we summarize results of a density matrix renormalization
group (DMRG) study \cite{white} of isotropic Heisenberg coupled-chain
systems with $n_c=1$, $2$, $3$, and $4$.
We find that the $n_c=2$ and $n_c=4$ systems have a spin gap, while
the $n_c=1$ and $n_c=3$ systems are gapless.
Furthermore,
Affleck's generalization\cite{afflecklsm} of the Lieb-Shultz-Mattis (LSM)
theorem for coupled chain systems shows that all odd-$n_c$ systems
are gapless.
Based on these results,
we discuss how a variational RVB wave function, originally introduced
by Liang et. al.\cite{liang} to describe the 2D antiferromagnetic
Heisenberg system, provides an intuitive picture
for understanding
the results and suggests behavior for larger $n_c$ than can be studied
numerically.

We consider the Heisenberg Hamiltonian
\begin{equation}
H = J \sum_{\langle i,j \rangle} {\hbox{\bf S}}_{i} \cdot {\hbox{\bf S}}_{j}
\end{equation}
defined on an $L\times n_c$ lattice with $S=\frac12$.
We will also consider the anisotropic system where the exchange along
the chains is $J$ and between the chains is $J'$, but unless otherwise
noted, $J'=J$.
We begin by calculating the spin gap $\Delta$ defined by
\begin{equation}
\Delta(L) = E_0(L,1) - E_0(L,0).
\end{equation}
Here $E_0(L,S_z)$ is the ground state energy for an $L\times n_c$ lattice
with open boundary conditions and
$z$-component of total spin $S_z$. For a single chain we expect
that the finite size corrections scale as $L^{-1}$, and we have
plotted $\Delta(L)$ versus $L^{-1}$ in Fig. 1 for $n_c=1$--$4$.
The solid curves are fits to the data of the form
\begin{equation}
\Delta(L) = \Delta + a_1 L^{-1} + a_2 L^{-2} + \ldots
\end{equation}
For $n_c=1$, we use $\Delta=0$, $a_1 = 4.09$, $a_2 = -11.2$.
For $n_c=2$, the data is fit very well with $a_1=0$, which is
the expected form for an $S=1$ Heisenberg chain \cite{sorensen}:
we find $\Delta = 0.5037$, $a_2 = 21.24$, $a_3 = -162.8$, $a_4 = 470.8$.
For $n_c=3$, we have $\Delta=0$, $a_1 = 2.61$, $a_2 = 4.11$.
For $n_c=4$, we have $\Delta=0.209$, $a_1 = 0$, $a_2 = 10.1$.
The spin-spin correlation functions
$\langle {\hbox{\bf S}}_i\cdot{\hbox{\bf S}}_j\rangle$
are shown in Fig. 2. Here,
because of the open boundary conditions, we have chosen $i$ and $j$ so
that they are as symmetrically located about the center of the lattice as
possible. The semlilog plot in the inset of Fig. 2(a) shows the exponential
decay of the spin correlations for the two and four chain systems.
The correlation
length for $n_c=2$ is $\xi=3.19(1)$, and for $n_c=4$, $\xi \sim 5-6$.
The spin-spin correlations for $n_c=3$ decay
as a power law, similar to those for a single ($n_c=1$) chain,
as shown in the inset of Fig. 2(b).

The LSM theorem states that a half-integer spin
chain, with a Hamiltonian that has local couplings and rotational and
translational symmetry, either has gapless excitations or else has degenerate
ground states. Affleck proved a similar statement
for coupled spin chains\cite{afflecklsm}: an isotropic coupled-chain
system with half-integer spin and
a finite, odd number of chains either has gapless excitations or else has
degenerate ground states.
For even-$n_c$ systems, the theorem does not apply.

Haldane's conjecture\cite{haldane} that
single Heisenberg spin chains containing integer spins have gaps, while
those containing half-integer spin do not,
has by now been fairly well established.
The inapplicability of the generalized LSM theorem for even-$n_c$
systems, and our
DMRG results for $n_c=2$ and $4$ (along with other $n_c=2$
results\cite{dagotto,strong,barnes,parola})
suggest that a similar statement may be true for coupled Heisenberg
$S=\frac12$ chain systems:
for odd $n_c$ the spin gap vanishes, while for even $n_c$ there is
a spin gap. To obtain a more intuitive picture, we examine an RVB
variational wavefunction\cite{anderson,kivelson}.
Previous descriptions of the spin-liquid state of the $n_c=2$
ladder have used a mean field approach which represents spins in
terms of bond operators for the rungs \cite{rice2}.
Here, in contrast, we consider variational RVB wavefunctions for
the ground state and conclude
that a short-range RVB picture
applies for even $n_c$, whereas a long-range RVB picture describes
systems with odd $n_c$.

The RVB states we consider are specific to bipartite lattices, and
contain only bonds connecting one sublattice ($A$) to the other ($B$).
We consider wave functions of the form \cite{liang}
\begin{equation}
|\psi\rangle = \sum_{i_\alpha \in A \atop j_\alpha \in B }
h(i_1-j_1) \ldots h(i_n-j_n)(i_1j_1) \ldots(i_nj_n),
\end{equation}
Here $(ij)$ represents a singlet bond between sites $i$ and $j$, and
the non-negative bond amplitude $h$ can be chosen variationally.
We consider a short-range RVB wavefunction to be one with a
bond amplitude $h(l)$
which decays exponentially in $l$ or faster, while a long-range
RVB wave function
will typically have a power-law decay, $h(l) \sim l^{-p}$.

For the $n_c=2$ system, we first consider a specific dimer RVB ansatz,
such that $h(i-j)=1$ for $i$ and $j$ nearest neighbors, and is
zero otherwise.
A variety of properties of the system are qualitatively, although
not quantitatively predicted by this simple variational state.
Unlike the equivalent two-dimensional system \cite{liang}, where
a Monte Carlo calculation is necessary,
for the ladder the properties of
the system can be calculated analytically.
However, its most appealing feature is that many of its most important
characteristics can
be understood without the need of detailed calculations.

A valence bond configuration for this state is formed by drawing dimer bonds
connecting pairs of adjacent sites, with every site part of one bond.
The resonance between different valence bond configurations leads to
a substantial lowering of the energy.
The simplest and perhaps most important type of resonance consists of
a square of four adjacent sites fluctuating between two adjacent
vertical bonds and two adjacent horizontal bonds\cite{kivelson}.
Consider the possible resonances for a ladder system.
The two types of bond configurations, ``resonating'' and
``staggered'', are shown in Fig. 3(a) and 3(b), respectively.
The staggered type of
configuration is incapable of resonance, and thus has higher
energy.  It is possible to form a local region of staggered bond order
only by placing soliton spin defects
at the edges of the region, as shown in Fig. 3(c).  Hence to
lowest order the staggered bond configurations can be ignored.  We
assume that all resonating configurations are equally likely, and
the ground state, within this variational estimate, is taken as the
sum of all such configurations.
The properties of this state can be obtained using the loop-covering
method developed by Sutherland \cite{sutherland,liang}.
We use a recursion relation for the normalization of $\psi$ as
a function of the ladder length $L$; from this, we can find spin-spin
correlation functions and the energy of the system.
The derivation will be published elsewhere.
One obtains an average energy per site of -0.556029.
Compared with the essentially exact result from the DMRG
calculations of -0.578043140, the simple dimer RVB energy differs by
less than 4\%.
While the variational energy is reasonable, the spin-spin correlation
length $\xi = 0.238012$ calculated with this dimer RVB state is more than
an order of magnitude smaller than our DMRG result of $\xi = 3.19$.
This implies that $h(l)$ has a larger range. However, as discussed for
the 2D lattice in Ref. \cite{liang}, as long as $h(l)$ falls off
exponentially  one finds an exponential decay of spin correlations
and a spin gap.

Although, as we have seen, the correlation length is poorly determined with
the dimer RVB ansatz, a variety of qualitative feautures
predicted by the ansatz are indeed present. For example, the variational
state has a greater bond strength for interchain nearest-neighbor
bonds compared to nearest-neighbor intrachain bonds. Most
importantly, within the short-range RVB picture one expects to find that
pairs of topological spin defects are bound. We
see from Fig. 3(c) that two spin defects produce a region of staggered
bond order between them if they are separated.
Furthermore, one expects from this picture that
the pair of defects should reside predominantly on a single rung, as
in Fig. 3(d), rather than on adjacent sites on a single chain,
in order to maximize resonance.
Each of these predictions is supported by
the DMRG calculations.

If we remove one of the sites of the lattice
from both the first and last rungs, as
shown in Fig. 3(b), in order to {\it force} the system to have staggered
bond order, we expect a topological spin defect to appear at each
end to remove the staggering effect. The resulting spin defects are confined
to the ends of the lattice, and are similar to the effective $S=\frac12$
spins on the ends of open $S=1$ chains \cite{hagiwara,white}.
As in that case, instead of an
isolated ground state, we have a singlet and a triplet of states with
a separation in energy which falls off exponentially with $L$.
Figure 4 shows DMRG results for the local spin moment and
nearest-neighbor bond strengths in
the vicinity of a modified end of an $n_c=2$ lattice which
contains one of these localized spin defects. Both the staggered bond
order and localized spin defect are clearly visible. The greater bond
strength for rung bonds in the center of the system is also apparent.

Now, it is possible to represent {\it any } singlet state as an RVB
state \cite{liang}, provided long-range singlet bonds are allowed.
(In order to represent any singlet state, Eq. (2) must be generalized
somewhat. Nevertheless, Eq. (2) has been shown to work extremely
well for the ground state of the 2D system \cite{liang}.)
In 2D, even when the probability amplitude of a long-range
bond decays as a power-law with the separation,
one can still obtain a finite staggered magnetization, provided
the power is sufficiently small \cite{liang}.
For a finite number of coupled chains, however, it seems more
likely that power law decay of bond amplitudes always gives power
law decay of spin-spin correlations.
The crucial point in considering such an RVB representation
is whether the amplitude for long-range bonds decays exponentially
or algebraically, and if algebraically, with what exponent. Our DMRG
results indicate that for the $n_c=2$ and $n_c=4$ systems
the universality class
is that of the short-range RVB, i.e. exponential decay of
bond amplitudes. For quantitative results from the variational state,
we expect that we must include some longer bonds and optimize
over the bond amplitudes, but for qualitative results, the dimer
state is adequate. The generalization of the LSM theorem, plus our results
for $n_c=3$, indicate that the universality class for odd $n_c$ is
the long-range RVB state.

What is the behavior for even $n_c > 4$, and why is there different
behavior for odd and even $n_c$? We believe the answer to this can be
understood in terms of the
confinement of topological defects present within a dimer RVB state with
even $n_c$. The confinement for $n_c=2$ is represented in Fig. 3(c), and
the lack of confinement for $n_c=3$ is shown in Fig. 3(e). In general,
for even $n_c$, the presence of a single defect puts the system into
a generalized form of staggered order characterized by an odd number
of bonds crossing any vertical line separating rungs.
We expect that this staggered
order, although still capable of resonance for $n_c \ge 4$,
is higher in energy than the ``resonating'' type of order.
Thus defects are
confined for an even number of chains,
just as for the $n_c=2$ case illustrated in Fig. 3(c)\cite{footone}.
For odd $n_c$, there is only one type of
order, characterized by an alternation as one moves along the chains
of an odd number and an even number of bonds crossing a vertical line.
A defect shifts the alternation by one lattice spacing, but with no
cost in energy away from the defect.

The confinement of defects relates to the presence of long-range bonds
in the ground state because a long-range bond can be considered
to be a pair of separated topological defects.
Thus considering a single long bond in a background of dimer bonds,
we expect ``confinement'' of the long bond for even $n_c$; in other
words, we expect it to be supressed exponentially with the separation,
since the energy difference grows linearly with the size of the
staggered region. (In making this argument, we are allowing the
region between the two sites connected by the long bond to resonate
between different valence bond configurations,
while holding the long bond fixed. The same conclusion is obtained if
we instead consider the {\it number } of valence bond configurations which
have such a long bond.)
Note also that the presence of non-dimer, but still
short-ranged bonds does not heal the staggered order induced by
the long-range bond. Such a short-range bond only heals the staggered
order within the region of the bond.
The presence of these
short-range non-dimer bonds can be considered as ``dressing''
the dimer state, lowering the energies of regions with resonant
bond order and with staggered order, but not changing the result
that the staggered-order region is higher in energy.
If a sufficiently high density
of non-dimer bonds were present, the confinement picture might
not be valid, but variational calculations for the 2D Heisenberg
model show that even in long-range, low-energy RVB states, dimer
bonds are much more probable than any other type of bond \cite{liang}.
Thus it appears that this confinement mechanism is very effective
at suppressing long-range bonds.

Since the characteristic size for this confinement mechanism is the
system width $n_c$, we expect that for even $n_c$, the
spin-spin correlation length varies as $\xi \sim n_c$,
corresponding to a spin gap varying as $1/n_c$.
For odd $n_c$, no confinement occurs, and the system is free to
have long-range bonds. Although this, in itself, does not show
that bond amplitudes decay as a power-law, both our
numerical results and the generalization of the LSM theorem
provide evidence that they do. The results suggest that in general,
unless there is some mechanism to suppress long-range bonds, such as
the confinement mechanism, we should expect power-law decay of
bond amplitudes.

The confinement argument applies also to the anisotropic case,
with $J' \ne J$. The most interesting case is $J' < J$.
{}From our RVB picture
we expect a gap to be present for all finite $J'/J$. As $J' \to 0$,
the number of vertical bonds decreases, and the difference in
energy between the staggered and resonant types of bond configurations
decreases. Nevertheless, the energy difference should be nonzero for
all finite $J'$, and confinement should cause exponential falloff
of the bond amplitudes $h(l)$, giving exponential spin-spin correlations
and a finite gap. We expect similar behavior for any system with an
even number of chains.
This prediction is in agreement with the conclusions for $n_c=2$ of
Strong and Millis \cite{strong} and the numerical evidence of
Barnes et. al. \cite{barnes}. In addition, using the DMRG method,
we have calculated the
gap as a function of system length $L$ for
$J'=0.1$, $J=1$, and $n_c=2$ for $L$ as large as 100. We find
a gap of 0.05988 for the infinite system.

For even $n_c$, we expect that the confinement mechanism applies also
to charge defects. In particular, for $n_c=2$
we expect that a single hole will consist primarily of an empty
site and a spin defect located on the same rung, in
agreement with the results of Tsunetsugu, et. al.\cite{tsunetsugu},
for the $t$-$J$ model.
Similarly, two holes will be bound, and will primarily consist of
two empty sites on the same rung.
For odd $n_c$, the lack of confinement of long-range bonds does not
necessarily imply spin-charge separation, although it does occur
for $n_c=1$.

The authors thank M. Bander for useful discussions.
S.R.W. and R.M.N. acknowledge support from the Office of Naval
Research under grant No. N00014-91-J-1143 and D.J.S. acknowledges support
from the National Science Foundation under grant DMR92--25027.
This work was supported  in part by the University of California
through an allocation of computer time.

\begin{figure}
\caption{Spin gaps as a function of system size $L$ for open $L\times n_c$
coupled chain Heisenberg systems.
}
\label{a}
\end{figure}
\begin{figure}
\caption{Spin-spin correlations $\langle{\hbox{\bf S}}_i\cdot{\hbox{\bf
S}}_j\rangle$ versus
$|i-j|$ with $i$ and $j$ located on the top chain
for (a) $n_c$ even. The semilog plot in the inset shows the
exponential decay of the correlations.
(b) $n_c$ odd. The log-log plot
in the inset shows that the correlations for $n_c=3$ and $n_c=1$ decay
with similar power-laws.
The deviation from pure power-law behavior visible for the largest values
of $|i-j|$ is due to finite-size effects from the open boundaries.}
\label{b}
\end{figure}
\begin{figure}
\caption{Various dimer valence bond configurations, with and without
topological spin defects present.
}
\label{c}
\end{figure}
\begin{figure}
\caption{One end of a long ($L=50$) $n_c=2$ chain with the first site
of the bottom chain missing. A topological spin defect
(an $S=\frac12$ up spin)
is trapped near the end of the chain. The defect ``heals'' the
staggered bond order imposed by the modified chain end. In (a) we
show the local magnetization, and in (b) we show the nearest-neighbor
bond strengths.
}
\label{d}
\end{figure}


\begin{references}
\bibitem{footone} If it turns out that the staggered order
is lower in energy, then the ground state has staggered order and
there is still binding of defects because of the induced resonating
order between the defects.
\end{references}
\end{document}